\documentclass[twocolumn]{aastex631}

\shorttitle{Protocluster cross-correlation function}
\shortauthors{Ramakrishnan et al.}
\graphicspath{{./}{figures/}}
\usepackage{natbib}
\usepackage{amsmath}
\usepackage{amssymb}
\usepackage{nicefrac}

\newcommand{\sqdeg}{\mbox{${\rm deg}^2$}}

\def\h50{\, h_{50}^{-1}}

\def\ltsima{$\; \buildrel < \over \sim \;$}
\def\simlt{\lower.5ex\hbox{\ltsima}}
\def\gtsima{$\; \buildrel > \over \sim \;$}
\def\simgt{\lower.5ex\hbox{\gtsima}} 
\def\arcsec{$''$}

\begin{document}

\title{ODIN: Strong Clustering of Protoclusters at Cosmic Noon}

\author[0000-0002-9176-7252]{Vandana Ramakrishnan}
\affiliation{Department of Physics and Astronomy, Purdue University, 525 Northwestern Avenue, West Lafayette, IN 47907, USA}

\author[0000-0003-3004-9596]{Kyoung-Soo Lee}
\affiliation{Department of Physics and Astronomy, Purdue University, 525 Northwestern Avenue, West Lafayette, IN 47907, USA}

\author[0000-0002-9811-2443]{Nicole Firestone}
\affiliation{Physics and Astronomy Department, Rutgers, The State University, Piscataway, NJ 08854}

\author[0000-0003-1530-8713]{Eric Gawiser}
\affiliation{Physics and Astronomy Department, Rutgers, The State University, Piscataway, NJ 08854}

\author[0000-0003-0570-785X]{Maria Celeste Artale}
\affiliation{Departamento de Ciencias Fisicas, Universidad Andres Bello, Fernandez Concha 700, Las Condes, Santiago, Chile}

\author[0000-0001-6842-2371]{Caryl Gronwall}
\affiliation{Department of Astronomy \& Astrophysics, The Pennsylvania
State University, University Park, PA 16802, USA}
\affiliation{Institute for Gravitation and the Cosmos, The Pennsylvania
State University, University Park, PA 16802, USA}

\author[0000-0002-4902-0075]{Lucia Guaita}
\affiliation{Departamento de Ciencias Fisicas, Universidad Andres Bello, Fernandez Concha 700, Las Condes, Santiago, Chile}

\author[0009-0003-9748-4194]{Sang Hyeok Im}
\affiliation{Department of Physics and Astronomy, Seoul National University, 1 Gwanak-ro, Gwanak-gu, Seoul 08826, Republic of Korea}

\author[0000-0002-2770-808X]{Woong-Seob Jeong}
\affiliation{Korea Astronomy and Space Science Institute, 776 Daedeokdae-ro, Yuseong-gu, Daejeon 34055, Republic of Korea}

\author[0009-0002-3931-6697]{Seongjae Kim}
\affiliation{Korea Astronomy and Space Science Institute, 776 Daedeokdae-ro, Yuseong-gu, Daejeon 34055, Republic of Korea}

\author[0000-0001-6270-3527]{Ankit Kumar}
\affiliation{Departamento de Ciencias Fisicas, Universidad Andres Bello, Fernandez Concha 700, Las Condes, Santiago, Chile}

\author[0000-0002-6810-1778]{Jaehyun Lee}
\affiliation{Korea Astronomy and Space Science Institute, 776 Daedeokdae-ro, Yuseong-gu, Daejeon 34055, Republic of Korea}%\affiliation{Korea Institute for Advanced Study, 85 Hoegi-ro, Dongdaemun-gu, Seoul 02455, Republic of Korea}

\author[0009-0008-4022-3870]{Byeongha Moon}
\affiliation{Korea Astronomy and Space Science Institute, 776 Daedeokdae-ro, Yuseong-gu, Daejeon 34055, Republic of Korea}

\author[0000-0001-9850-9419]{Nelson Padilla}
\affiliation{Instituto de Astronomía Teórica y Experimental (IATE), CONICET-UNC, Laprida 854, X500BGR, Córdoba, Argentina}

\author[0000-0001-9521-6397]{Changbom Park}
\affiliation{Korea Institute for Advanced Study, 85 Hoegi-ro, Dongdaemun-gu, Seoul 02455, Republic of Korea}

\author[0000-0002-4362-4070]{Hyunmi Song}
\affiliation{Department of Astronomy and Space Science, Chungnam National University, 99 Daehak-ro, Yuseong-gu, Daejeon, 34134, Republic of Korea}

\author[0000-0001-6162-3023]{Paulina Troncoso}
\affiliation{Escuela de Ingeniería, Universidad Central de Chile, Avenida Francisco de Aguirre 0405, 171-0614 La Serena, Coquimbo, Chile}

\author[0000-0003-3078-2763]{Yujin Yang}
\affiliation{Korea Astronomy and Space Science Institute, 776 Daedeokdae-ro, Yuseong-gu, Daejeon 34055, Republic of Korea}

%% Mark off the abstract in the ``abstract'' environment. 
\begin{abstract}
The One-hundred-deg$^2$ DECam Imaging in Narrowbands (ODIN) survey is carrying out a systematic search for protoclusters during Cosmic Noon, using Ly$\alpha$-emitting galaxies (LAEs) as tracers. Once completed, ODIN aims to identify hundreds of protoclusters at redshifts of 2.4, 3.1, and 4.5 across seven extragalactic fields, covering a total area of up to 91~deg$^2$. In this work, we report strong clustering of high-redshift protoclusters through the protocluster-LAE cross-correlation function measurements of 150 protocluster candidates at $z~=~2.4$ and 3.1, identified in two ODIN fields with a total area of 13.9~deg$^2$. At $z~=~2.4$ and 3.1, respectively, the inferred protocluster biases are $6.6^{+1.3}_{-1.1}$ and $6.1^{+1.3}_{-1.1}$, corresponding to mean halo masses of $\log \langle M /M_\odot\rangle = 13.53^{+0.21}_{-0.24}$ and $12.96^{+0.28}_{-0.33}$. By the present day, these protoclusters are expected to evolve into virialized galaxy clusters with a mean mass of $\sim$ $10^{14.5}~M_\odot$. By comparing the observed number density of protoclusters to that of halos with the measured clustering strength, we find that our sample is highly complete. Finally, the similar descendant masses derived for our samples at $z=2.4$ and 3.1 assuming that the halo number density remains constant suggest that they represent similar structures observed at different cosmic epochs. As a consequence, any observed differences between the two samples can be understood as redshift evolution. The ODIN protocluster samples will thus provide valuable insights into the cosmic evolution of cluster galaxies.

\end{abstract}

\section{Introduction} \label{sec:intro}

The effect of environment on the formation and evolution of galaxies is a question of considerable importance. In the local Universe, clusters of galaxies display an excess of massive and quiescent galaxies compared to the coeval field \citep[e.g.,][]{Peng2010,Quadri2012}. These galaxies are expected to have assembled the bulk of their mass at high redshifts ($z~\sim2-4$). Indeed at $z~\gtrsim~2$, the progenitors of present-day clusters of galaxies (`protoclusters') host copious star-formation and AGN activity \citep[e.g.,][]{Casey2015,Oteo2018,Harikane2019,Staab2024} as well as evolved galaxies \citep{Ito2023,Jin2024}, suggesting ongoing rapid mass assembly. Thus, in order to understand the growth of galaxies in overdense regions, it is necessary to study protoclusters.

In recent years, studies have begun identifying increasingly large samples of protoclusters through various metods \citep[e.g.,][]{Chiang2014,Toshikawa2016,Toshikawa2018,Higuchi2019}. Yet, connecting these objects with clusters identified at lower redshifts remains challenging. This is in large part because protoclusters remain unvirialized and lack several of the characteristics of clusters such as a red sequence or hot intra-cluster medium \citep[see][for a review]{Overzier2016}. As they are observed in the initial phases of their formation, it is difficult to determine what characteristics they may ultimately exhibit.

One of the protocluster properties most useful for judging their evolution is the halo mass, as we expect that the halo mass of the most massive halo within a protocluster is indicative of the mass of the final cluster \citep{Chiang2013}. One way to determine the halo mass is through measurements of the clustering strength of the protoclusters, as more massive halos are also more strongly clustered \citep{Sheth1999}. However, such measurements require a large sample to carry out, which has prevented them from being widely used so far. 

Recently, we presented the selection of protoclusters from the One-hundred-deg$^2$ DECam Imaging in Narrowbands \citep[ODIN,][]{Lee2024} survey in \citet{Ramakrishnan2024}. By imaging a wide area of the sky with three custom narrowband filters, ODIN has made it possible to identify a large and robust protocluster sample at Cosmic Noon. In this work, we explore the clustering of ODIN protoclusters selected at $z~=~2.4$ and 3.1. In Section \ref{sec:data}, we present our protocluster sample. We analyze the clustering using the two-point correlation function in Section~\ref{sec:cross-corr}. We interpret our results for the halo mass and probable descendants of the ODIN structures in Section \ref{sec:discussion}. Finally, we summarize our findings in Section~\ref{sec:summary}. Throughout, we assume a Planck cosmology \citep{Planck2016} with $\Omega_{\rm \Lambda} = 0.6911$, $\Omega_{\rm b} = 0.0486$, $\Omega_{\rm m} = 0.3089$, $H_{0} = 100\,h\,{\rm km}\,{\rm s}^{-1}\,{\rm Mpc}^{-1}$ and $h=0.6774$. Distances are given in comoving units, with an implicit $h_{68}^{-1}$.

\section{Protocluster samples} \label{sec:data}

\begin{figure*}
    \centering
    \includegraphics[width=0.95\linewidth]{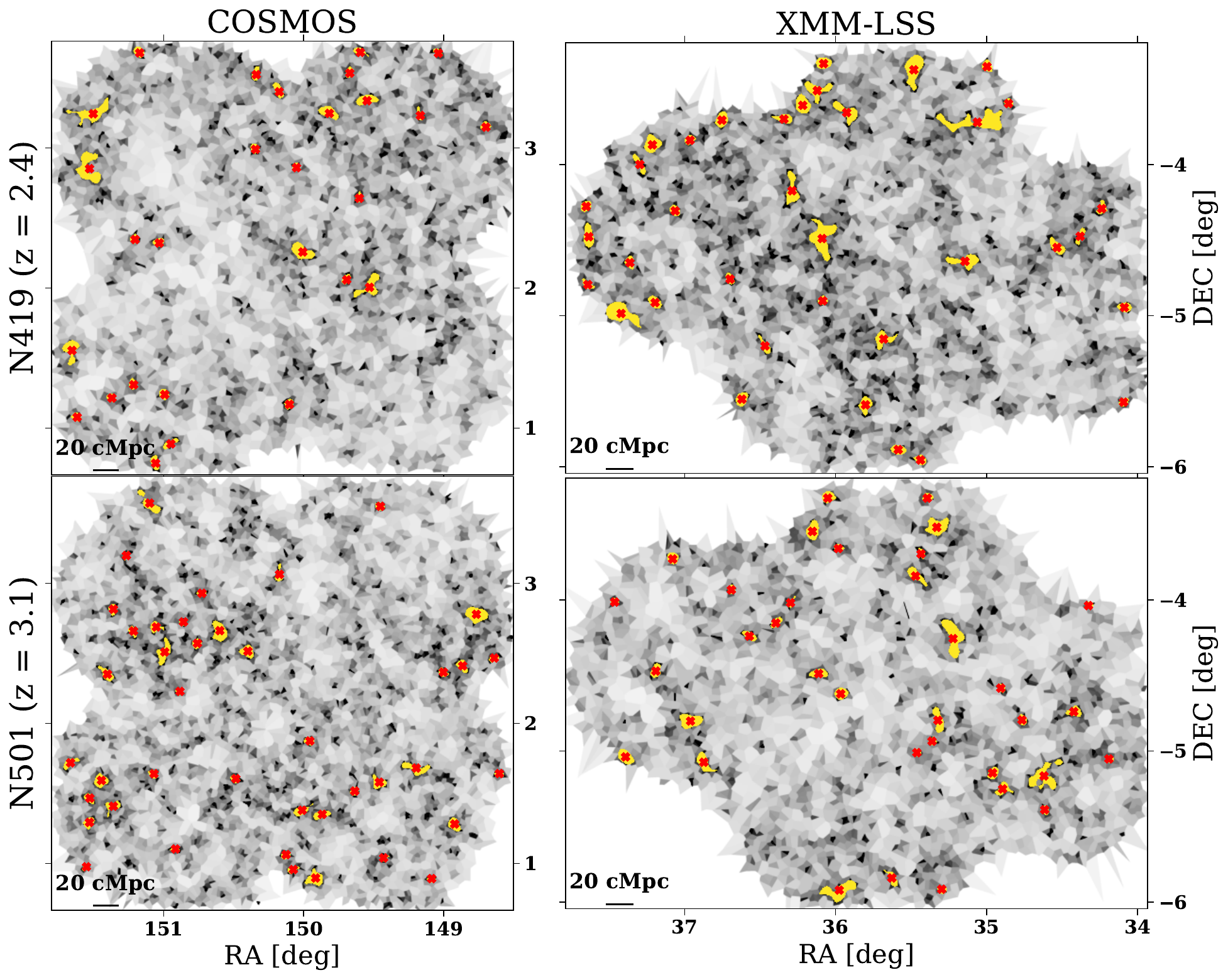}
    \caption{Protocluster candidates (yellow swathes) selected with our 4 field/filter combinations, overlaid on the Voronoi Tessellation-based LAE surface density maps. The shape of the density maps is a result of the arrangement of the four Deep HSC-SSP pointings in each field. The geometric centers of each protocluster are indicated by red crosses. The density threshold and minimum used to select these objects are given in the text.}
    \label{fig:pc_candidates}
\end{figure*}

The ODIN survey is a wide-field imaging survey spanning 91~deg$^2$ of the equatorial and southern skies, utilizing three narrow-band filters—$N419$, $N501$, and $N673$—designed to detect Ly$\alpha$ emission at redshifts $z=2.4$, 3.1, and 4.5, respectively. As described in the ODIN survey paper \citep{Lee2024}, one of its primary objectives is to map large-scale structures using Ly$\alpha$-emitting galaxies (LAEs), a subset of low-mass, star-forming galaxies \citep[e.g.,][]{Gawiser2006,Guaita2010}. Previous angular clustering measurements suggest that LAEs reside in moderate-mass halos with low bias \citep[e.g.,][]{Gawiser2007, Guaita2010, Lee2014, Kusakabe2018, Hong2019, White2024}.

\begin{deluxetable*}{cccccccc}
    \tablecaption{Key parameters of large protocluster samples \label{tab:data}}
    \tablehead{\colhead{Redshift} & \colhead{Field} & \colhead{Effective area} & \colhead{V$_{\rm eff}$} & \colhead{NB depth\tablenotemark{a}} & \colhead{$N_g$\tablenotemark{b}} & \colhead{$N_p$\tablenotemark{c}} & \colhead{$n_p$} \\
    \colhead{} & \colhead{} & \colhead{[\sqdeg]} &\colhead{[10$^6$ cMpc$^3$]} & \colhead{} & \colhead{} & \colhead{} & \colhead{[10$^{-6}$ cMpc$^{-3}$]}}
    \startdata
        $2.45\pm 0.03$ & COSMOS & 7.3 & 4.6 & 25.37 & 6,441 & 28 & 6.1 \\
        (ODIN) & XMM-LSS & 6.6 & 4.2 & 25.25 & 5,614 & 36 & 8.6 \\
         & Total & 13.9 & 8.8 & & {\bf 12,055} & {\bf 64} & {\bf 7.3} \\
        \hline
        $3.12\pm 0.03$ & COSMOS & 7.3 & 5.8 & 25.37 &  6,069 & 41 & 7.0 \\
        (ODIN) & XMM-LSS & 6.6 & 5.2 & 25.47 & 3,928 & 35 & 6.6 \\
         & Total & 13.9 & 11.0 & & {\bf 9,997} & {\bf 76} & {\bf 6.9} \\
        \hline
        3.8 $\pm$ 0.45\tablenotemark{d} & HSC-SSP Wide & 121 & 1,510 & & 366,690\tablenotemark{e} & 216 & 0.14 \\
        %COSMOS & $N419$ & 25.0 & 6441 & 28 \\ 
        %XMM-LSS & $N419$ & 25.0 & 5614 & 36 \\ 
        %COSMOS & $N501$ & 25.0  & 6069 & 41 \\ 
        %XMM-LSS & $N501$ & 25.0 & 3928 & 35 \\ 
        %COSMOS & N673 & 25.0 & 4233 & 25 \\ 
        %XMM-LSS & N673 & 25.0 & 3231 & 28 \\ 
    \enddata
    \tablenotetext{a}{Median 5$\sigma$ depth measured in a 2\arcsec\ diameter circular aperture}
    \tablenotetext{b}{Number of LAEs \citep[see][]{Firestone2024}}
    \tablenotetext{c}{Number of protocluster candidates \citep[see][]{Ramakrishnan2024}}
    \tablenotetext{d}{Key numbers of the protocluster search conducted by \citet{Toshikawa2018} using LBGs as tracers}
    \tablenotetext{e}{From \citet{Ono2018}}
\end{deluxetable*}

The photometric selection of ODIN LAEs combines the ODIN narrow-band (NB) imaging data with deep broad-band imaging from the Hyper Suprime-Cam Subaru Strategic Program \citep[HSC-SSP;][]{Aihara2018a, Aihara2018b}. The selection method is detailed in \citet{Firestone2024} and is summarized as follows: (i) LAEs must be detected in the NB image with a signal-to-noise ratio of at least 5, and (ii) they must exhibit excess flux in the NB compared to the estimated continuum flux density near the Ly$\alpha$ wavelength. The NB-to-BB color cut corresponds to a rest-frame equivalent width of 20~\AA. Sources near bright stars are excluded using HSC-SSP star masks \citep{Coupon2018}. The effective area of each field, after excluding the star-masked regions, and the number of detected LAEs are provided in Table~\ref{tab:data}.

The selection method for the protoclusters and the expected contamination rate are detailed in \citet{Ramakrishnan2024}. Briefly, protoclusters are identified from maps of the LAE surface density ($1 +\delta_{\rm LAE}$) as contiguous regions that exceed predefined thresholds in both density and angular size ($A$). These criteria are optimized to minimize false detections while maximizing the total number of protoclusters. We follow the procedure described in \citet{Ramakrishnan2024} to estimate the contamination rate and adopt the threshold of $\delta_{\rm LAE}=2.3$ (2.6), corresponding to a significance of 4$\sigma$ (4.5$\sigma$) above the field surface density, for our protocluster selection at $z=2.4$ (3.1). The lower detection threshold for the $z~=~2.4$ sample is possible because of the higher LAE number density at that redshift.
 For both redshifts, the minimum area is chosen to be $A=40$~cMpc$^2$.
%and $A=40$ (40) ~cMpc$^2$  and (2.6, 40 cMpc$^2$) for the $N419$ and $N501$ filters respectively. 

A potential concern in ensuring the robustness of our protocluster detection is the variation in imaging sensitivity across the narrow-band data used for LAE selection. ODIN employs DECam, which covers a circular area of $2.1^\circ$ in diameter, to image a larger field matching the LSST field of view (FOV) of $3.5^\circ$. Although a wide dither scheme is used to ensure a reasonable level of uniformity within the LSST FOV \citep[see Figure~6 and Section~4.1 in][]{Lee2024}, the central region of $1^\circ$ in radius is approximately 5\% shallower in limiting flux than the average depth. Additionally, edge effects near the image boundaries where photometric noise is greater could masquerade as LAE overdensities. To address this, we apply a magnitude cut to the LAE sample when constructing the surface density map. Multiple realizations of `random' density maps are also created by randomly selecting NB-detected sources (\emph{without} any significant narrowband excess), and overdensity regions flagged in these maps are visually inspected. We find that the overdensities seen in the random maps do not significantly overlap with the regions where real protoclusters are detected, demonstrating that our detected protoclusters are unlikely to arise from noise fluctuations.

Figure~\ref{fig:pc_candidates} shows the locations of ODIN protoclusters as yellow swaths, with red crosses marking each protocluster’s geometric center. The final count of protoclusters in each field is provided in Table~\ref{tab:data}. The effective volume, $V_{\rm eff}$, is calculated by converting the filter transmission curve into the redshift selection function. For comparison, the table also includes data from the protocluster sample used for clustering measurements in \citet{Toshikawa2018}. Their sample, based on Lyman break galaxy (LBG) overdensities, features a much wider redshift selection function and covers an area approximately ten times larger than the ODIN sample. Table~\ref{tab:data} shows that the number density of LBG-detected protoclusters is about 50 times lower than that of ODIN protoclusters, suggesting that the LBG sample may trace rarer, more massive systems (see later). 

%We select the protocluster candidates in a similar manner to the COSMOS/$N501$ sample for the remaining field/filter combinations. We apply the same selection criteria consistently across both fields for each filter. The resulting protocluster candidates overlaid on the corresponding surface density maps, are shown in Figure~\ref{fig:pc_candidates}. Further details of the protocluster selection and resulting candidates for all our field/filter combinations (including the $N673$ filter) will be discussed in an upcoming work (V. Ramakrishnan et al.\ in prep.).

In our earlier work \citep{Ramakrishnan2024}, we utilized data from IllustrisTNG, a suite of cosmological hydrodynamical simulations, to better understand the selection efficiency of ODIN protoclusters and to calibrate estimates of their descendant masses based on the level and angular extent of the LAE overdensity ($\delta_{\rm LAE}$ and $A$). We created mock ODIN volumes from the $z~=~3$ snapshot, centered on cosmic structures selected by their present-day mass. In doing so, we closely match the ODIN survey parameters including the line-of-sight thickness and the surface density of galaxies. For the latter, we randomly selected a subset of galaxies within a specific mass range to approximately match the observed clustering strength of LAEs \citep[e.g.,][]{White2024}. The mock data were then projected onto the XY, YZ, or XZ planes to create multiple sky images where we applied the protocluster detection criteria. For protoclusters with descendant masses of $M_{z=0}\gtrsim 2\times 10^{14}M_\odot$, the recovery rate is $\approx$60\%. Non-detection of about 40\% occurs because, along some sightlines, TNG structures did not satisfy both criteria for the detection. Using the mass calibration, we find that the real ODIN protoclusters have a median descendant mass of $M_{\rm z=0}\sim 10^{14.5}M_\odot$.

\section{Protocluster cross-correlation measurements} \label{sec:cross-corr}

The two-point correlation function (CF) is a straightforward method for measuring the clustering strength of a set of points. When applied to the positions of astronomical sources, the CF measure can be compared with the analytical or numerical models for the clustering of dark matter halos \citep[e.g.,][]{mo1996}, thereby providing insight into the properties of cosmic structures hosting them \citep[e.g.,][]{zehavi04,lee06,harikane22}. Similarly, the presence or absence of an evolutionary connection between two samples observed at different cosmic epochs can be established based on the CF, as the redshift evolution of halo clustering is known. 

Because protoclusters are unvirialized structures, they are composed of multiple dark matter halos. The canonical theory of hierarchical structure formation dictates that all halos will merge with \emph{the most massive halo within the protocluster} and become substructures within the dynamical timescale. In this context, the clustering measurements at scales larger than the size of a protocluster probe the properties of these most massive halos, though for brevity, we will refer to quantities such as `protocluster bias' in the following discussion. It is worth stating that, in the subsequent sections, we focus on the clustering of these parent halos and their redshift evolution, and not on internal clustering within a given protocluster structure.

\subsection{Protocluster-LAE cross-correlation function}

The most direct method for estimating the bias and mass of the most massive dark matter halos within the protoclusters is to measure the two-point auto-correlation function, as done by \citet{Toshikawa2018}. However, our sample at each redshift is about one-third the size of theirs, leading to significant uncertainties in the auto-correlation function measurement. Instead, we opt to measure the cross-correlation function (CCF) between protoclusters and LAEs. Because the clustering properties of LAEs are well understood both in general \citep[e.g.,][]{kovac07,Guaita2010,Lee2014} and for our samples \citep[][D. Herrera et al., in prep]{White2024}, this approach helps reduce measurement uncertainties, provided that certain assumptions hold (discussed later). As the ODIN survey nears completion, full auto-correlation function measurements will be presented in a future paper.

\begin{figure*}[t!]
    \centering
\includegraphics[width=0.85\linewidth]{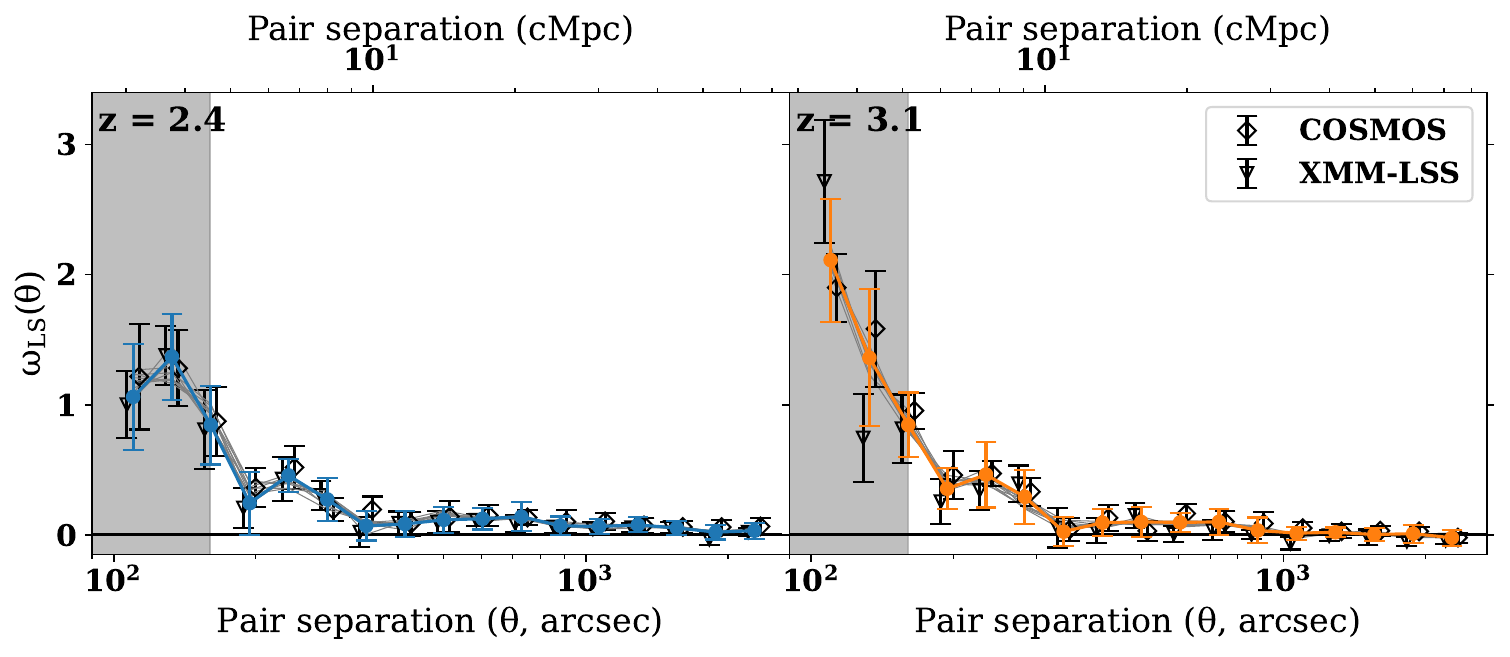}
    \caption{The cross-correlation function measurements are shown as a function of angular separation for $z=2.4$ (left) and $z=3.1$ (right), with the corresponding projected comoving separations displayed on the top axes. In each panel, the combined CCF is represented by color-filled circles and thick solid lines of the same color, while the individual measurements for the COSMOS and XMM-LSS fields are shown as black diamonds and downward triangles, respectively (offset slightly for clarity). Thin grey lines show the combined measurements using 10 random catalogs. The grey-shaded regions mark the median angular size of the ODIN protoclusters. A strong signal is detected at both redshift ranges.
%    from our protocluster and LAE samples, shown for the COSMOS and XMM-LSS fields individually (open symbols) and for the combined samples (filled symbols). A robust signal is detected at both $z~=~2.4$ and $z~=~3.1$. 
}
    \label{fig:cross_corrs}
\end{figure*}

Using the \citet{LandySzalay1993} estimator, we calculate the two-point protocluster-LAE cross-correlation function, $\omega_{\rm LS}(\theta)$:
\begin{equation} 
\omega_{\rm LS}(\theta)=\frac{D_{p} D_{g} (\theta) -D_{p} R_{g}(\theta)-D_{g} R_{p} (\theta)+ R_{p} R_{g}(\theta)}{R_{g}R_{p}(\theta)}
%\omega_{\rm LS}(\theta)=\frac{\rm D_{g}D_{p}-D_{g}R_{p}-D_{p}R_{g}+R_{g}R_{p}}{\rm R_{g}R_{p}}
\label{eq:eq1}
\end{equation}
where $D_pD_g(\theta)$ is the number of protocluster-LAE pairs at angular separation of $\theta$, with `p' and `g' denoting protoclusters and LAEs, respectively. $R_pR_g(\theta)$ is the analogous quantity for random pairs, and $D_pR_g(\theta)$ and $D_gR_p(\theta)$ are the number of random-source cross pairs. Logarithmic binning of the angular separations is chosen to ensure adequate sampling at both small- and large scales. 

The position of each protocluster is shown as red crosses in Figure~\ref{fig:pc_candidates}, overlaid on the LAE density map for each field. As described in \citet{Ramakrishnan2024}, the local LAE density is measured and converted into a pixelated map with an angular resolution of 3\farcs6. A source detection algorithm is applied to this map to identify protoclusters and flag the pixels belonging to each structure. The position of each protocluster is defined as its geometric center, calculated as the geometric mean of all the associated pixels. Our main results remain consistent even if we use the pixel with the peak LAE density as the protocluster center.

To construct a random galaxy catalog ($R_g$), we select 20,000 sources with a signal-to-noise ratio greater than 5 from the narrow-band catalog such that their magnitude distribution matches the LAEs. This approach allows us to account for small depth variations that might affect our LAE selection by applying the same selection effect to our random sources. Ten independent random catalogs are created for use. As for protocluster randoms ($R_p$), we generate 20,000 random points within the field, as the protocluster distribution only depends on the relative LAE \emph{overdensity} and thus is insensitive to the survey depth. The same star masks used for our LAE sample are applied to both protoclusters and our random catalogs. 

%We first randomly select a subset of 20,000 NB-detected sources with SNR $>$ 5, such that their NB magnitude distribution matches that of the LAEs. This is necessary to capture the effect of any depth variations on the LAE surface density. For the protoclusters, we use instead a set of 20,000 uniformly distributed random points, following \citet{Toshikawa2018}; this is reasonable since the distribution of the protoclusters is independent of the survey depth. We apply the same star masks used for the LAE sample to both the protocluster and LAE randoms. 

Each pair count in Equation~\ref{eq:eq1} is made separately in the two fields and then combined as:
%We calculate the combined and normalized pair counts from the two fields as, 
\begin{equation}
    X_p X_g(\theta_j) = \frac{\sum_i X_{p,i} X_{g,i}(\theta_j)}{\sum_i\sum_kX_{p,i} X_{g,i}(\theta_k)}
\label{eq:eq2}
\end{equation}
where $X$ is either $D$ (data) or $R$ (random), $i$ refers to the $i$th field, and $j$  or $k$ indexes the angular bins. The covariance matrix, $\mathcal{C}$, for the CCF is computed using jackknife sampling by dividing the samples \emph{in each field} into 15 subsamples and repeating the measurement 15 times, each time leaving out one subsample. ${\mathcal C}$ has the dimension of $N_\theta \times N_\theta$ where $N_\theta$ is the number of angular bins in our CCF measures. The choice of 15 subsamples was determined by varying the number and identifying the minimum number at which the covariance matrix converged.

The resulting CCFs at both redshifts are shown in Figure~\ref{fig:cross_corrs} where the same measures for the COSMOS and XMM-LSS fields are also shown. In each case, the error bars represent the square root of the diagonal of the covariance matrix. For both redshift samples, the individual measures are consistent with the combined function within the uncertainties. 
%This suggests that our selection is picking up on similar objects in both fields.

\subsection{Protocluster bias}\label{subsec:proto_bias}

%The protocluster-LAE cross-correlation function, $\omega_X$, is related to the underlying matter correlation function, $\omega_m$, by the bias $b_X$,
The bias $b_X$ determines the scaling between the protocluster-LAE CCF, $\omega_X$, and matter correlation function, $\omega_m$, as:
\begin{equation}\label{eqn:bias}
    \omega_X(\theta) = b_X^2 \omega_m(\theta)
\end{equation}
where the former is related to our observational measure through the integral constraint (${\mathcal IC}$), which accounts for the loss of power at scales larger than our survey fields and thus depends on the size of the field and the galaxy and protocluster power spectrum. 
%The function $1+\omega(\theta)$ represents the total probability of finding one object at a separation $theta$ from another, and thus when integrated over to infinite separations should be 1. On the other hand the integral of $1+\omega_{LS}$ over the finite area of our survey is also 1. That is, $\omega_{LS}(\theta)$ is offset from $\omega(\theta)$, with the two being related by an integral constraint, $IC$. 
The two are related as:
\begin{align}\label{eqn:LS_model}
    1 + \omega_{\rm LS}(\theta) &= \frac{1 + \omega_X(\theta)}{1 + {\mathcal IC}} 
%    \omega_{\rm LS}(\theta) &= \frac{\omega(\theta) - IC}{1+ IC}
\end{align}
Following the prescription given by \citet{Adelberger2005} \citep[also see,][]{Infante1994,Roche1999}, we estimate ${\mathcal IC}$ as, 
\begin{equation}\label{eqn:IC}
    {\mathcal IC}=\frac{\sum_j {R_g R_p(\theta_j)} \omega_X(\theta_j)}{\sum_j {R_g R_p(\theta_j)}} = b_X^2 \frac{\sum_j {R_g R_p (\theta_j)} \omega_m(\theta_j)}{\sum_j {R_g R_p (\theta_j)}}
\end{equation}
%Thus from Equation \ref{eqn:bias} we have,
By combining Equations~\ref{eqn:bias}, \ref{eqn:LS_model} and \ref{eqn:IC}, we can write down $\omega_{\rm LS}(\theta)$ in terms of all other quantities as:
\begin{align}\label{eqn:LS_bias_model}
%    {\rm IC} &= b_X^2 \frac{\sum {\rm RR} \omega_m(\theta)}{\sum {\rm R_gR_p}} \\ \nonumber
%    \omega_{\rm LS}(\theta) &= b_X^2\frac{\omega_m(\theta) - \nicefrac{\sum {\rm R_gR_p} \omega_m(\theta)}{\sum {\rm R_gR_p}}}{1+b_X^2\nicefrac{\sum {\rm R_gR_p} \omega_m(\theta)}{\sum {\rm R_gR_p}}}
\omega_{\rm LS}(\theta) = b_X^2 \frac{\omega_m(\theta)-[\sum_j R_g R_p (\theta_j) \omega_m (\theta_j)/\sum_j R_g R_p (\theta_j)]}{1+b_X^2[\sum_j R_g R_p (\theta_j) \omega_m (\theta_j)/\sum_j R_g R_p (\theta_j)]}
\end{align}
We calculate $b_X$ using this equation where $\omega_m(\theta)$ is obtained from the angular matter power spectrum using the Core Cosmology Library \citep[CCL,][]{ccl_paper}.

The uncertainty in $b_X$ is estimated by generating 5,000 realizations of the CCF according to a multivariate normal distribution of the form exp[$(x_i-\mu_i){\mathcal C}^{-1}_{ij}(x_j-\mu_j)$]. Here, $x_i$ is the value of the perturbed CCF in the angular bin $\theta_i$ and $\mu_i$ and ${\mathcal C}_{ij}$ are the measured CCF, $\omega_{\rm LS}(\theta_i)$, and its covariance, respectively. Each realization is fit using Equation~\ref{eqn:LS_bias_model} to determine the corresponding $b_X$ value. For the fit, we only include the angular bins with $\theta\geq 160$\arcsec, which corresponds to a projected separation of 4.6 (5.0)~cMpc at $z=2.4$ (3.1), roughly matching the typical size of the ODIN protoclusters\footnote{The size is computed as the median effective radius as $\sqrt{A/\pi}$ where $A$ is the transverse area rising above the adopted detection threshold $\delta_{\rm LAE}$ for a given protocluster. See Section~\ref{sec:data} for more details.}. At scales smaller than a protocluster, the CCF is expected to deviate from a power-law due to LAEs residing in the same protocluster (the ``one-halo'' term). As illustrated in Figure~\ref{fig:bias_fits}, the best-fit power-laws to our data lie significantly below the measured values for angular separations of $\theta\lesssim 160$\arcsec. 
The exclusion of the small-scale clustering from the fitting means that our results are insensitive to the uncertainty in protocluster centers. In Table~\ref{tab:measurements}, we report the best-fit $b_X$ values where the upper and lower limits indicate the 16th and 84th percentile, respectively.

%are the measured cross-correlation and its covariance. We fit each realization with Equation \ref{eqn:LS_bias_model} to find the distribution of $b_X$. We do not fit the bins corresponding to a separation of less than 160\arcsec, corresponding to a physical separation of 4.6~cMpc at $z~=~2.4$ and 5~cMpc at $z~=~3.1$. The value of 160\arcsec\ arises from the median effective radius of our protoclusters, which is calculated based on the area $A$ above the density threshold (see Section \ref{sec:data}) as $\sqrt{A/\pi}$. 

%This lower limit ensures that our results are insensitive to the uncertainty in the centroid of the protocluster, and also excludes the region of the CCF which is dominated by the correlation of a protocluster with the LAEs which lie withiin its boundaries and are used to define it. The resulting values and uncertainty of $b_X$, found as the median and 16-84 percentile of the distribution, are reported in Table~\ref{tab:measurements}. The best-fitting versions of Equation~\ref{eqn:LS_bias_model}) (using the median bias values) are shown in Figure~\ref{fig:bias_fits}.

The CCF bias, $b_X$, is the geometric mean of the protocluster bias ($b_p$) and LAE bias ($b_g$). The LAE bias is known from the clustering measurements of ODIN LAEs \citep[][D. Herrera et al., in prep]{White2024}, which yield in $1.7 \pm 0.2$ at $z=2.4$ and $2.0 \pm 0.2$ at $z=3.1$. Using this, we can now estimate the protocluster bias as $b_{p} = {b_{X}^2}/{b_{g}}$. 
%The value of $b_X$ inferred above is the geometric mean of the biases of the protocluster and LAE samples, $b_{PC}$ and $b_{g}$. Thus if the LAE bias at each redshift is known, we can infer the protocluster bias as,
%\begin{equation}
%    b_{PC} = \frac{b_{X}^2}{b_{g}}
%\end{equation}
%We use the values of $b_{g}$ found by \citet{White2024}, who measured the clustering of LAEs chosen independently of the ODIN selection from the $N419$ and $N501$ data in the COSMOS field and confirmed with DESI. 
The results are summarized in Table~\ref{tab:measurements} where the uncertainties in both $b_X$ and $b_g$ are propagated into those in $b_p$. The inferred protocluster bias is $b_p\approx 6-7$, much higher than any known galaxy population. Previous studies showed that LBGs at similar redshift typically have $b\sim 2-3$ \citep{lee06,hildebrandt09,harikane22} while more UV-faint LAEs tend to have even lower biases, $b\sim 1.5-2.0$ \citep[e.g.,][]{Guaita2010,White2024}. The high bias values imply that ODIN-selected structures correspond to more massive dark matter halos than these galaxy populations.

%The resulting bias values for our protocluster candidates (Table \ref{tab:measurements}) are much higher than the LAE biases and those of LBGs at a similar redshift. This indicates that the ODIN structures correspond to halos which are much more massive than individual galaxies. %With these values, the biases of our protocluster candidates are $b_{PC}$ = 5.7$^{+1.7}_{-1.5}$ and 6.1$^{+1.4}_{-1.2}$. These values are much higher than the LAE biases and those of LBGs \textcolor{blue}{cite}.

\begin{deluxetable*}{cccccccccc}
    \tablecaption{Biases, correlation lengths, and halo masses of ODIN protoclusters \label{tab:measurements}}
    \tablehead{\colhead{Redshift} & \colhead{$b_X$} & \colhead{$b_g$\tablenotemark{a}} & \colhead{$b_p$\tablenotemark{b}} & \colhead{$r_{0,X}$} & \colhead{$r_{0,g}$\tablenotemark{a}} & \colhead{$r_{0,p}$\tablenotemark{b}} & \colhead{log($M_{\rm min}/M_\odot$)} & \colhead{log $\langle M \rangle/M_\odot$ } & \colhead{$n_h$ } \\ 
    & & \colhead{(fixed)} & & \colhead{(cMpc)} & \colhead{(fixed, cMpc)} & \colhead{(cMpc)} & & & \colhead{(10$^{-6}$ cMpc$^{-3}$)}}
    \startdata
    2.4 & 3.36$^{+0.21}_{-0.23}$ & 1.7 $\pm$ 0.2 & 6.6$^{+1.3}_{-1.1}$ & 10.10$^{+0.82}_{-0.84}$ & 4.3 $\pm$ 0.3 & 23.7$^{+4.2}_{-4.3}$ & 13.31$^{+0.24}_{-0.26}$ & 13.53$^{+0.21}_{-0.24}$ & 2.8$^{+9.5}_{-2.2}$\\
    3.1 & 3.47$^{+0.33}_{-0.33}$ & 2.0 $\pm$ 0.2 & 6.1$^{+1.3}_{-1.3}$ & 8.78$^{+1.01}_{-0.99}$ & 4.3 $\pm$ 0.3 & 17.9$^{+4.3}_{-4.2}$ & 12.76$^{+0.32}_{-0.39}$ & 12.96$^{+0.28}_{-0.33}$ & 15.5$^{+8.3}_{-1.3}$\\
    \enddata    
    \tablenotetext{a}{ Measured LAE bias from \citet{White2024}}
    \tablenotetext{b}{Uncertainties are estimated by propagating the errors on input quantities assuming that they are uncorrelated.}
\end{deluxetable*}

%We calculate the LAE autocorrelation function from the two fields and find $b_{g}$ analogously to the protocluster-LAE cross-correlation function. We find that $b_{g}$ $=$ ---. These values are consistent with the results of Herrera et al.\ (in prep).\ which make use of only the LAE samples in the COSMOS field. With these LAE biases, the resultant protocluster biases are ----.

\subsection{Protocluster correlation length}

\begin{figure*}[th!]
    \centering
    \includegraphics[width=0.85\linewidth]{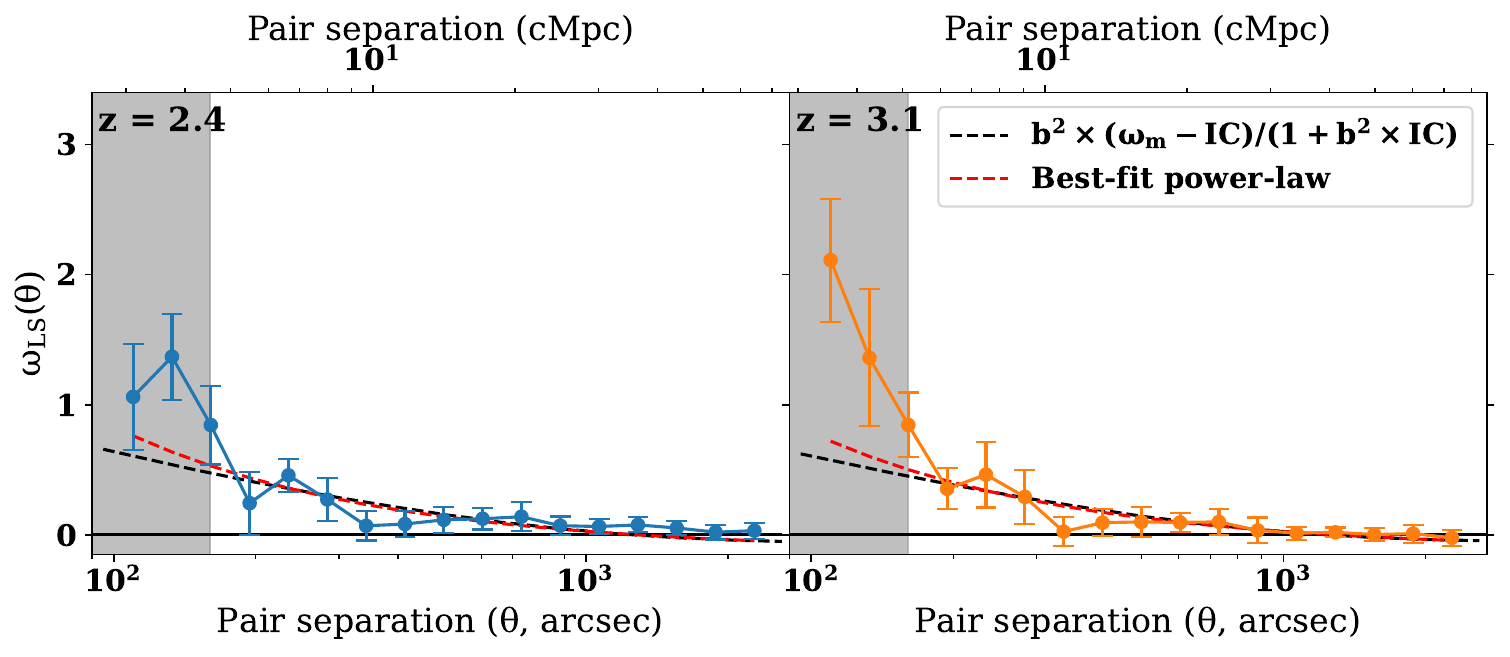}
    \caption{Cross-correlation functions together with median bias values (black dashed lines) which give the best fit of Equation \ref{eqn:LS_bias_model} to our observed cross-correlation functions. Red dashed lines show the best-fit power-law model with integral constraint.}
    \label{fig:bias_fits}
\end{figure*}

If the real-space correlation function follows a power law, $\xi(r) = (r/r_0)^{-\gamma}$, where $r_0$ is the correlation length, the corresponding angular correlation function, $\omega(\theta)$, takes the form $A_{\omega} \theta^{-\beta}$ and is related to $\xi(r)$ through the \citet{Limber1953} equation: \begin{equation}
\label{eq:Limber} 
A_{\omega} = C r_{0,X}^{\gamma} \int_{0}^{\infty} F(z) D_\theta^{1-\gamma}(z) N(z)^2 g(z) dz, 
\end{equation} 
where $\gamma = 1 + \beta$, $F(z)$ describes the redshift dependence of $\xi(r)$, $D_\theta(z)$ is the angular diameter distance at redshift $z$, $N(z)$ is the redshift selection function for the galaxy sample, and \begin{equation}
\label{g(z)} 
g(z) = \frac{H_0}{c} \left[ (1+z)^2 \left( 1 + \Omega_M z + \Omega_\Lambda \left[ (1+z)^{-2} - 1 \right] \right)^{\frac{1}{2}} \right]. 
\end{equation} 
Here, $C = \sqrt{\pi}\Gamma((\gamma - 1)/2)/\Gamma(\gamma/2)$. Given the narrow transmission widths of our filters (corresponding to a line-of-sight thickness of 60--70 cMpc), it is reasonable to assume $F(z) = 1$. We convert the total throughput of each filter to obtain $N(z)$ \citep[see Figure 1 of][]{Lee2024}.

Under the power-law assumption, we estimate the correlation lengths as follows. As before, we generate 5,000 realizations of the CCF from our $\omega_{\rm LS}$ measurements. Using Equations~\ref{eqn:LS_model} and \ref{eqn:IC}, we determine the best-fit $A_\omega$. The $\beta$ value is fixed to 0.8 to match the result from \citet{White2024}; it is also consistent with the power-law slopes found or assumed by existing studies. The correlation length, $r_{0,X}$, is calculated using Equation~\ref{eq:Limber}. Re-fitting the CCF after fixing $\beta$ to 0.6 or 1.0 yield functionally identical results for the correlation length, which is unsurprising given that $\beta$ and $A_\omega$ are strongly correlated. The best-fit power-law with the median value of $A_{\omega}$ in Figure~\ref{fig:bias_fits} together with our CCF measures. 

If the real-space correlation functions for both protoclusters and LAEs share the same power-law slope, $\gamma$, it can be shown that their correlation lengths are related through:
\begin{equation} \label{eqn:r0}
    r_{0,p} = \frac{r_{0,X}^2}{r_{0,g}}
\end{equation}
We adopt $r_{0,g}=4.3\pm 0.3$ cMpc for both redshift samples as reported by \citet{White2024}. Table~\ref{tab:measurements} lists the best-fit $r_{0,X}$ and $r_{0,p}$ values. 

As expected from the large bias values we obtained earlier, the inferred correlation lengths for ODIN protoclusters are large, $r_{0,p}=18-24$~cMpc. For comparison, for flux-limited (${\mathcal R} \leq 25.5$) LBG samples at $z\sim 2.2$ and 2.9, \citet{Adelberger2005} reported $r_0=4.2\pm 0.5$ and $4.0\pm 0.6$~$h^{-1}$~cMpc, respectively. For more UV-luminous LBGs ($r \leq 24.5$) at $z\sim 3$, \citet{hildebrandt09} found $r_0=5.9\pm 0.4$~$h^{-1}$~cMpc, in agreement with luminosity-dependent clustering strength \citep[e.g.,][]{gd01,lee06,harikane22}. Finally, we note that, at $z\sim 3.8$, \citet{Toshikawa2018} estimated the correlation length for their LBG-selected protocluster candidates to be $35.7^{+4.6}_{-5.5}$~$h^{-1}$~cMpc, i.e., much larger than our value. In Section~\ref{subsec:halo_mass}, we discuss this difference in the context of redshift evolution of the abundance and clustering properties of halos.

\section{Discussion}\label{sec:discussion}

\subsection{Halo mass and abundance of ODIN protoclusters}\label{subsec:halo_mass}

Both the space density and clustering strength of halos strongly depend on halo mass, with more massive halos being rarer and more biased tracers of the underlying matter distribution compared to less massive ones. In this section, we use the analytic framework developed by \citet{Sheth1999} and \citet{Mo_and_White2002} to analyze our data and infer the properties of the halos hosting ODIN protoclusters. All calculations are performed using the CCL library routines.

If the minimum mass of the halos associated with an ODIN protocluster is $M_{\rm min}$, the average bias of the sample can be expressed as:
\begin{equation} \label{eqn:M_min}
\langle b_h \rangle = \frac{\int_{M_{\rm min}}^{\infty} b(M) n(M) dM}{\int_{M_{\rm min}}^{\infty} n(M) dM}
\end{equation} 
where $b(M)$ is the bias of halos with mass $M$, and $n(M)$ is the halo mass function, following the \citet{Sheth1999} model. The best-fit $M_{\rm min}$ value is determined by setting $b_p = \langle b_h \rangle$, where $b_p$ is the protocluster bias obtained in Section~\ref{subsec:proto_bias}. Once $M_{\rm min}$ is known, the mean halo mass $\langle M \rangle$ is calculated as:
\begin{equation} \label{eqn:M_mean}
    \langle M \rangle = \frac{\int_{M_{\rm min}}^{\infty} M n(M) dM}{\int_{M_{\rm min}}^{\infty} n(M) dM}.
\end{equation}

The values of $M_{\rm min}$ and $\langle M \rangle$, listed in Table~\ref{tab:measurements}, suggest that the most massive halo within the ODIN protoclusters has a typical mass of $\langle M \rangle \approx 10^{13} M_\odot$ at the time of observation. The similarity between the  $M_{\rm min}$ and $\langle M \rangle$ values is expected, as the steep slope of the halo mass function means that the number of halos in a sample is dominated by those close to the cut-off mass.

Finally, we calculate the number density of halos whose clustering matches the observed value by integrating the halo mass function above $M_{\rm min}$:
\begin{equation}
    n_h = \int_{M_{\rm min}}^{\infty} n(M) dM
\end{equation}
By comparing this to the number density of protoclusters, $n_p$, the ratio $n_p/n_h$ gives an estimate of the completeness of the protocluster selection. For example, if only 50\% of halos above a given mass threshold are observationally detected as a protocluster, both samples would have the same clustering strength, but $n_p/n_h = 0.5$. 

While $n_p/n_h > 1$ is theoretically unphysical, sample contamination or uncertainties in the measured protocluster bias could lead to ratios exceeding unity. As structures observed well before virialization, protoclusters are expected to be extended and have irregular morphologies. The sky distribution of galaxies in several well-studied protoclusters supports this view \citep[e.g.,][]{hayashino04,Dey2016,Cucciati2018}.  Depending on the viewing angle, a single structure may be fragmented into more than one structure, leading to an overestimation of $n_p$. Conversely, the opposite effect could also occur, where $n_p$ is underestimated.

%As protoclusters are highly extended and irregular in shape, making it difficult to clearly determine their extent, there is also a chance that a single structure may be fragmented into more than one object during the detection. This would also lead to $n_p$ being overestimated, although this should be offset in some measure by the opposite effect (i.e. multiple protoclusters being blended and detected as a single structure).

\begin{figure}
    \centering
    \includegraphics[width=0.9\linewidth]{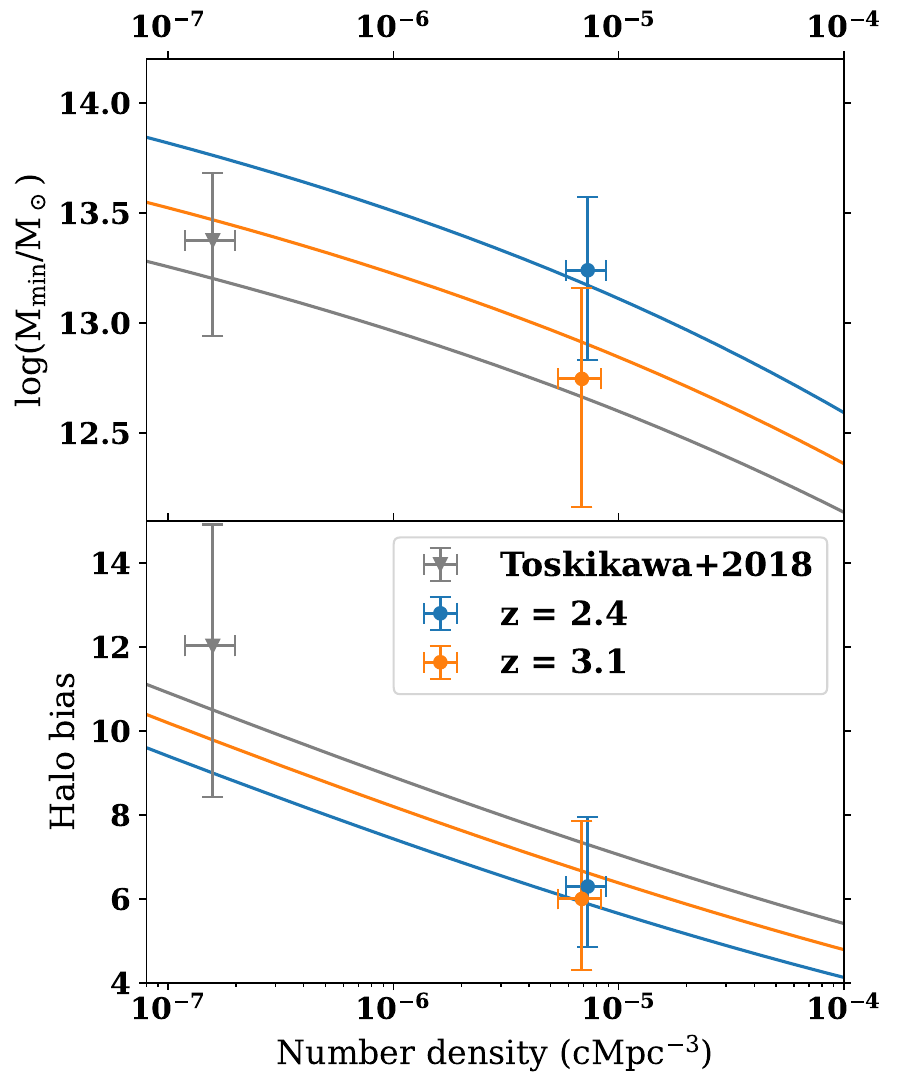}
    \caption{
    Color lines show the expected relationship between $M_{\rm min}$ (top) or $\langle b_h \rangle$ (bottom) and the number density of halos above the threshold ($n_h$) at a redshift of $z=2.4$ (blue), 3.1 (orange), and 3.8 (grey), representing the LAE-selected ODIN protoclusters and LBG-selected sample from \citet{Toshikawa2018}, respectively. The data points show the protocluster number density $n_p$ and the $M_{\rm min}$ and $b_p$ values inferred from clustering measurements. 
%    Minimum halo mass ($M_{min}$) and protocluster halo bias ($b_p$) versus number density. Blue (orange) lines indicate the expected values at $z~=~2.4$ (3.1). The halo mass and halo bias (inferred from the reported halo mass) of \citet{Toshikawa2018} are shown for comparison.
    }
    \label{fig:completeness}
\end{figure}

Using the data from Tables~\ref{tab:data} and \ref{tab:measurements}, we find $n_p/n_h = 2.7^{+18.6}_{-2.3}$ at $z = 2.4$ and $0.44^{+4.6}_{-0.40}$ at $z = 3.1$. The latter value is consistent with an independent estimate from mock datasets based on IllustrisTNG, as described in Section~\ref{sec:data} \citep[also see][for more detail]{Ramakrishnan2024}. Overall, we observe a high completeness (50–100\%) for ODIN protocluster selection. In contrast, the protoclusters identified by \citet{Toshikawa2018} show a significantly higher ratio of $n_p/n_h \approx 5.3$.

Our findings are visualized in Figure~\ref{fig:completeness}. In the top panel, the colored lines represent the expected relationship between $M_{\rm min}$ and $n_h$ for halos at the redshift of the ODIN protocluster sample and the \citet{Toshikawa2018} sample. Similarly, the lines in the bottom panel show the predicted $\langle b_h \rangle$-$n_h$ scaling relations. The data points correspond to the $b_p$ or $M_{\rm min}$ values inferred from clustering measurements plotted against $n_p$. By requiring that the halo and protocluster samples have the same clustering strength (i.e., identical $b_p$ or $M_{\rm min}$ values), the completeness, represented by the ratio $n_p/n_h$, can be determined.

%Using the halo mass function of \citet{Sheth1999} implemented in CCL, we find an expected number density of 2.8$^{+9.5}_{-2.2}$ $\times$ 10$^{-6}$~cMpc$^{-3}$ halos above the inferred halo mass threshold at $z~=~2.4$ and 15.5$^{+8.3}_{-1.3}$ $\times$ 10$^{-6}$~cMpc$^{-3}$ at $z~=~3.1$. By comparison, the number density of our \emph{observed} protocluster candidates are 7.3 $\times$ 10$^{-6}$~cMpc$^{-3}$ at $z~=~2.4$ and 6.9 $\times$ 10$^{-6}$~cMpc$^{-3}$ at $z~=~3.1$ (Table~\ref{tab:data}). This suggests that the completeness of our protocluster sample at $z~=~3.1$ is $\sim$ 45\% and that at $z~=~2.4$ is $\sim$ 100\%. The former is consistent with the prediction made in \citet{Ramakrishnan2024} (Section~\ref{sec:data}).

The protocluster correlation length and predicted halo mass in our sample are both lower than those reported by \citet{Toshikawa2018}—$51.7^{+4.4}_{-4.9}$~cMpc and $(3.4 \pm 0.7) \times 10^{13} M_\odot$ at $z \sim 3.8$—suggesting that their sample likely contains more massive structures. Given the near-unity completeness of our sample, this difference is unlikely to stem from fundamental differences in how LBGs and LAEs trace cosmic structures. Instead, it is likely a selection effect. The redshift range for LBG selection is much broader (with a half-width-at-half-maximum of $\Delta z \approx 0.3-0.4$) than for LAE-based selections like ODIN ($\Delta z \sim 0.03$), making LBG samples more susceptible to projection effects from foreground and background interlopers. As a result, the cosmic structures identified as significant LBG overdensities are likely to be more massive than those selected based on LAE overdensities.

%they select protoclusters as overdensities of LBGs. Because the redshift uncertainty of LBGs is much larger than that of our LAEs ($\Delta z~\sim~0.5$ compared to $\Delta z~\sim~0.03$), foreground and background contamination will likely reduce the significance of any density peaks \citep{Chiang2013}. Thus, only the most massive objects are likely to be sufficiently overdense to be identified.

\subsection{Projected evolution and likely descendants}

What kinds of structures will ODIN protoclusters evolve into by the present-day universe? In this section, we explore this question by examining the cosmic evolution of dark matter halos. The simplest method is to trace halos with a fixed number density across cosmic time, assuming that each parent halo associated with a protocluster at high redshift will evolve into a distinct halo hosting a galaxy cluster at $z=0$. Starting from the protocluster number density $n_p$ , we find that the mean descendant mass is  $\langle M_{z=0} \rangle \sim 3.5\times 10^{14}M_\odot$ and $3.6\times 10^{14}M_\odot$ at $z=2.4$ and 3.1, respectively. Possible caveats of this approach include selection bias and completeness. 
  
A similar approach, but using the $n_h$  value, should be more robust against counting errors, as the estimate is based on clustering. The inferred descendant masses are consistent within uncertainties, with $\langle M \rangle = 5.5^{+7.1}_{-3.6}~\times~10^{14}~M_\odot$ and $2.3^{+4.9}_{-1.8}~\times~10^{14}~M_\odot$ at $z=2.4$ and 3.1, respectively. ODIN protoclusters are expected to evolve into intermediate-mass clusters, similar to the Virgo cluster. The clustering-based masses are consistent with a heuristic estimate presented in \citet{Ramakrishnan2024}.

%Assuming that the number density of the halos hosting our protocluster candidates remains constant (i.e. that each protocluster evolves into a distinct cluster), we can infer their minimum halo mass $M_{min}$ and expected mass $\langle M \rangle$ at $z~=~0$ from the halo mass function. Based on the \emph{observed} number densities, the expected descendant mass of the $z~=~3.1$ protocluster sample is %$\gtrsim$ 1.8 $\times$ 10$^{14}$ M$_\odot$ $\langle M \rangle$ $\sim$ $3.6~\times~10^{14}~M_\odot$. That of the $z~=~2.4$ protocluster sample is very similar, %$\gtrsim$ 1.7 $\times$ 10$^{14}$ M$_\odot$ $\langle M \rangle$ $\sim$ $3.5~\times~10^{14}~M_\odot$. 
%On the other hand, a more accurate estimate based on the number density obtained from the clustering measurements ($n_h$, see Table~\ref{tab:measurements}) yields descendant masses of $\langle M \rangle$ = $2.3^{+4.9}_{-1.8}~\times~10^{14}~M_\odot$ and $5.5^{+7.1}_{-3.6}~\times~10^{14}~M_\odot$ for the $z~=~3.1$ and 2.4 samples, respectively. The difference between the two estimates is small, again demonstrating the high completeness of our sample. Thus, our protocluster candidates are likely to evolve into intermediate-mass clusters, consistent with the descendant mass estimate of log$(M_{z=0}/M_\odot)~=~14.5$ for the $z~=~3.1$ protoclusters in \citet{Ramakrishnan2024}.

\begin{figure}
    \centering
    \includegraphics[width=\linewidth]{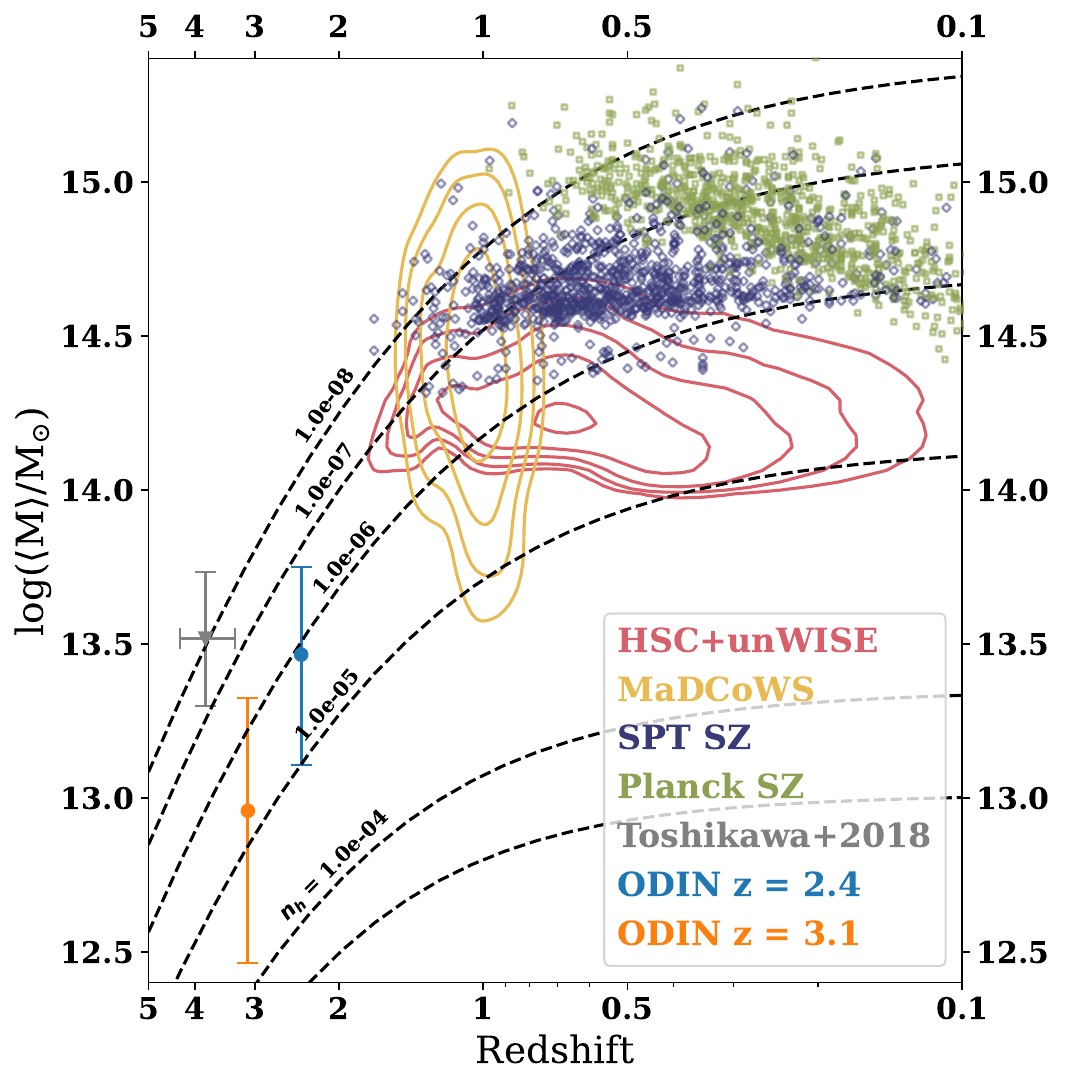}
    \caption{Mean halo mass of the ODIN protoclusters compared to various cluster samples at low redshift \citep[adapted from][]{Alberts2022}. Black dashed lines show the expected evolution of halo mass with redshift assuming a constant number density. %ODIN protoclusters are likely to evolve into structures less massive than SZ  clusters but similar to those selected with photometric redshifts.
An average ODIN protocluster will evolve into structures with masses comparable to high-redshift clusters identified through photometric selection.
    }
    \label{fig:halo_evolution}
\end{figure}

Figure~\ref{fig:halo_evolution} shows the redshift evolution of $\langle M \rangle $ at several fixed number densities. The dashed lines represent the mass growth history of a halo, assuming it does not merge with another halo of comparable mass.  Overlaid on the figure are observational measurements from recent cluster surveys compiled by \citet{Alberts2022}. These include clusters identified via the Sunyaev-Zeldovich (SZ) effect by {\it Planck} \citep{planck_SZ} and the South Pole Telescope \citep{bleem15,bleem20,huang20}, as well as clusters detected as high galaxy overdensities in the Massive and Distant Clusters of WISE Survey \citep[MaDCoWS:][]{Gonzalez2019} and a photometric redshift catalog based on Hyper-SuprimeCam (HSC) and unWISE data \citep{wen21}.

%, we plot the values of $\langle M \rangle$ inferred from the bias and track their expected evolution with redshift, assuming again that the number density stays constant. We also show the halo masses of various cluster samples identified at $z~\lesssim~1.5$. 
The figure illustrates that an average ODIN protocluster will evolve into structures with masses comparable to high-redshift  ($z=0.7-1.0$) clusters identified through photometric selection (HSC+unWISE in Figure~\ref{fig:halo_evolution}). The top 5--15\% of ODIN protoclusters\footnote{
This assumes that the completeness of ODIN protocluster detection is independent of descendant mass. However, in \citet{Ramakrishnan2024}, we demonstrate that more massive protoclusters are more likely to be detected. In a simple model where the completeness is 50\% below and 100\% above $M_{z=0}=10^{15}~M_\odot$, our estimate changes to 10-30\%. } will evolve into clusters with masses exceeding $10^{15}M_\odot$, significantly overlapping with MaDCoWS clusters at $z\approx 1$ and SZ-selected clusters at lower redshift. 
%
%The majority of our structures are likely to evolve into objects comparable to photometrically selected clusters at low redshifts, but are likely to be less massive than clusters selected based on the Sunyaev-Zeldovich effect. 
In contrast, the protoclusters studied by \citet{Toshikawa2018} are expected to evolve into some of the rarest and most massive clusters, far exceeding the mass of the Coma cluster. This is consistent with their much lower space density and higher minimum mass, even though they are observed at an earlier epoch ($z\approx 3.8$) compared to the ODIN protoclusters examined in this study (Figure~\ref{fig:completeness}).

%(which as noted previously appear to be intrinsically more massive objects) are probable progenitors of these more massive clusters with a well-developed hot intra-cluster medium.
%\textcolor{blue}{Based on the values of $M_{min}$ inferred at $z~=~0$, $\sim$ 5 - 15\% of our protoclusters are likely to evolve into the most massive structures with masses above 10$^{15}$ M$_\odot$. However, this is under the assumption that the likelihood of detecting a protocluster is independent of its descendant mass, i.e., the descendant masses of our observed protoclusters are distributed according to the halo mass function at $z~=~0$. In practice, we show in \citet{Ramakrishnan2024} that the protoclusters with the highest descendant mass are more likely to be successfully recovered. With an extremely simplistic model, wherein the recovery rate of a protocluster is 50\% for structures with $M_{z=0}~<~10^{15}~M_\odot$ and 100\% for the remainder, $\sim$ 10-30\% of our protoclusters are likely to represent these more massive objects.}

Finally, we highlight that the two ODIN protocluster samples have similar descendant masses, suggesting that they are likely comparable structures observed at different stages in their evolution. As a result, the results from ODIN will not only provide valuable insight into the properties of massive structures but also will enable us to directly trace the evolutionary sequence of galaxies therein over cosmic time. 

\section{Summary} \label{sec:summary}

%In this work, we seek to understand the properties of the halos hosting protoclusters at $z$ = 2.4 and 3.1, selected with data from the ODIN survey. 
%We use a sample of $\sim$ 75 protoclusters at each redshift, selected as overdensities of LAEs in the COSMOS and XMM-LSS fields (Table~\ref{tab:data}). 

We report strong clustering of protoclusters at $z=2.4$ and 3.1 identified as significant LAE overdensities by the ODIN survey over a total area of $\approx$14~deg$^2$. Through cross-correlation function measurements, we infer the correlation length and halo bias of ODIN protoclusters. Both quantities are significantly higher than those typically observed for galaxy populations at these epochs, validating that ODIN protocluster selection indeed pinpoints the sites of the rarest, most massive halos. By comparing the protocluster number density to that of halos with similar clustering amplitude,  we estimate the completeness of ODIN protocluster selection to be of the order of unity, consistent with an independent estimate from cosmological hydrodynamical simulations \citep{Ramakrishnan2024}.

At $z=2.4$ and 3.1, the mean masses of the halos hosting these protoclusters are $\log \langle M /M_\odot\rangle = 13.53^{+0.21}_{-0.24}$ and $12.96^{+0.28}_{-0.33}$, respectively. The top 10–30\% of ODIN protoclusters will evolve into the most massive clusters ($M_{z=0}\gtrsim 10^{15}M_\odot$), similar to SZ-selected clusters at low redshift, while the average ODIN protocluster is likely to become a more typical, lower-mass cluster detected by photometric surveys. The range of descendant masses for ODIN protoclusters is in good agreement with more heuristic estimates based on the level and the angular extent of the LAE overdensity calibrated against cosmological simulations. 
 
At both redshifts, the mean descendant masses ($M_{z=0}  \approx 10^{14.0-14.5}M_\odot$) are comparable, indicating that these protoclusters represent similar cosmic structures observed at different epochs. This close evolutionary relationship between the two samples suggests that any observed redshift-dependent trends can be reasonably interpreted as redshift evolution. The final ODIN protocluster samples are expected to be several times larger than those presented in this work and will extend out to $z~=~4.5$. ODIN will provide clean, well-calibrated samples of protoclusters that can be directly linked to lower-redshift systems, thereby elucidating the formation history of the present-day clusters of galaxies.

\begin{acknowledgments}
The authors acknowledge financial support from the National Science Foundation under Grant Nos. AST-2206705, AST-2408359, and AST-2206222, and from the Ross-Lynn Purdue Research Foundations. 
This material is based upon work supported by the National Science Foundation Graduate Research Fellowship Program under Grant No. DGE-2233066 to NF.
J.L. is supported by the National Research Foundation of Korea (NRF-2021R1C1C2011626).
H.S. acknowledges the support of the National Research Foundation of Korea (NRF) grant funded by the Korean government (MSIT) (No. 2022R1A4A3031306).
The Institute for Gravitation and the Cosmos is supported by the Eberly College of Science and the Office of the Senior Vice President for Research at the Pennsylvania State University.
Based on observations at Cerro Tololo Inter-American Observatory, NSF’s NOIRLab (Prop. ID 2020B-0201; PI: K.-S. Lee), which is managed by the Association of Universities for Research in Astronomy under a cooperative agreement with the National Science Foundation.
\end{acknowledgments}

\facility{Blanco}
\software{SExtractor \citep{Bertin1996}, SEP \citep{Barbary2016}, CCL \citep{ccl_paper}, Astropy \citep{astropy:2013,astropy:2018,astropy:2022}}

\bibliography{refs}{}
\bibliographystyle{aasjournal}

\end{document}